\documentstyle[twocolumn,floats,ifthen,aps,prb]{revtex}

\newcommand\eps{\varepsilon}
\newcommand\WW{\widetilde{W}}
\newcommand\DD{\widetilde{D}}

\begin{document}
\draft
%\preprint{Submitted to ...}

\title{Lateral photocurrent spreading in single quantum well infrared
photodetectors}

\author{M. Ershov}
\address{Department of Computer Software, University of Aizu, Aizu-Wakamatsu
  965-8580, Japan}

\date{February 4, 1998}
\maketitle

\begin{abstract}
Lateral physical effects in single quantum well infrared photodetectors (SQWIPs)
under non-uniform illumination over detector area are considered. These effects
are due mainly to the in-plane transport of the photoinduced charge in the QW.
The length of the lateral photocurrent spreading is determined by the in-plane
conductivity of the carriers in the QW and characteristic time of the QW
recharging, and can be as large as 10$^1$--10$^4$~$\mu$m. Closed-form analytical
expressions for SQWIP responsivity for modulated infrared signal and modulation
transfer function are obtained. Possible techniques to suppress lateral
photocurrent spreading are discussed.
\end{abstract}


%\newpage
\narrowtext
\vspace{1ex}

Quantum well infrared photodetectors (QWIPs) attracted a great deal of attention
in the last decade.~\cite{LevineR} Recently a number of new steady-state and
transient physical effects in QWIPs have been predicted theoretically and
confirmed
experimentally.~\cite{ErshovAPL95,ErshovAPLTr,ErshovNonlAPL,ErshovCapAPL}
Practically all theoretical studies of QWIPs assumed that infrared radiation
(IR) intensity is constant over the QWIP area, and that the physical processes
are uniform in the plane of the QWs. In practical conditions, however, IR
radiation intensity inside QWIP can be laterally non-uniform due to the
non-uniformity of the incoming infrared signal, light diffraction, reflection,
scattering and interference in QWIP, and other reasons. The non-uniformity of
the photoexcitation from the QWs and related lateral effects can change the
mechanism of operation and characteristics of QWIPs. The lateral effects are
especially critical for large-area pixelless QWIP-LED infrared imaging
devices.~\cite{Pixelless1,Pixelless2} The lateral spreading of the photocurrent
in QWIP determine the spatial resolution of the transformed infrared image, and
hence the imaging functionality of the device. One of the possible reasons for
image smearing in QWIPs with large photocurrent gain is the lateral spreading of
the photoinduced charge in the QWs near the emitter and resulting broadened
current injection from the emitter.~\cite{ErshovTransf}

In this letter we study theoretically the lateral photocurrent spreading and
related physical effects in QWIP with single QW (SQWIP). This case allows
analytical treatment due to its simplicity and thus leads to a better
understanding of the photocurrent spreading effects in QWIPs. Also, studying
SQWIPs will provide a worst-case estimate of the lateral spreading in QWIPs with
multiple QWs. In this letter we show that the lateral photocurrent spreading in
SQWIP is determined by the photoinduced charge screening and transport in the
QW, and, in particular, by the in-plane conductivity of the 2D electron gas
(2DEG) in the QW. The scale of the lateral spreading can be extremely large
(10$^1$--10$^4$~$\mu$m).

The SQWIP structure under consideration is similar to that considered
earlier~\cite{RyzhiiSJAP,ErshovIEEE96,ErshovAPLN} and is typical for practical
SQWIPs.~\cite{BandaraST93,BandaraS93} In short, SQWIP contains a narrow
single-level QW separated by wide undoped barriers from the contacts. For
concreteness, we consider $n$-type SQWIP. Electron concentration in the QW is
high (~10$^{11}$--10$^{12}$~cm$^{-2}$). Under applied voltage, electrons
injected from the emitter can be collected by the collector or captured into the
QW. The dynamics of the QW recharging is determined by the balance between the
capture of injected electrons and electron escape from the QW.

Let us first consider qualitatively the physical processes in SQWIP under
non-uniform illumination conditions (see Fig.~1). We are interested in the
lateral distribution of the photocurrent at the SQWIP output (collector
contact). Narrow infrared beam incident on SQWIP excites locally electrons from
the QW to the states above the barriers. These electrons reach collector or
emitter contact at the same lateral position (due to negligible out-diffusion)
and contribute to the photocurrent (primary photocurrent). Originally localized
photoinduced charge in the QW induce a lateral electric field, which causes
transport of the 2DEG in the QW towards the excitation area. Due to the high
2DEG conductivity, the resulting area of QW space charge becomes wide, inducing
injection from the emitter over a wide area. In the case of the large
photocurrent gain, most part of the injection current reach the collector,
resulting in significant spreading of the output photocurrent. Thus, the main
physical mechanisms of the lateral photocurrent spreading in SQWIP are the
lateral transport of the 2DEG in the QW and resulting spreading of the
photoinduced charge in the QW and injection from the emitter.

To quantitatively describe the SQWIP operation under non-uniform illumination
conditions, we use a 2D model of SQWIP which generalizes the 1D
model~\cite{RyzhiiSJAP,ErshovIEEE96,ErshovAPLN} for the case of lateral (along
the $y$-axis) non-uniformities of physical quantities. We use the small-signal
approximation, assuming that the photoemission from the QW due to the
non-uniform illumination is much weaker than the total emission due to thermo-
or photo-excitation. The sheet electron density in the QW $\delta \Sigma$ is
governed by the continuity equation (in the drift-diffusion approximation):

\begin{eqnarray} \label{cont-eq}
e\frac{\partial \delta\Sigma}{\partial t} +
\frac{d}{dy} \left[ e\mu \Sigma \frac{d\delta\varphi}{d y}
- e D \frac{d\delta \Sigma}{d y} \right] = \nonumber \\
(1-\beta)\, \delta j_e(E_e) -e\,\sigma\Sigma\, \delta I ,
\end{eqnarray}

\noindent where $\delta \varphi$ is the QW potential, $j_e$ is the injection
current density from the emitter dependent on emitter electric field $E_e$,
$\beta$ is the probability of the electron transport from emitter to collector,
$\sigma$ is the photoionization cross-section, $\mu$ and $D$ are the mobility
and diffusion coefficient of the 2DEG, and $\delta I$ is the infrared radiation
intensity. All small-signal quantities denoted by the $\delta$-symbol depend on
time and the $y$-coordinate. In this work we are interested in the case of the
long-range lateral non-uniformities with the scale much larger than the
thickness of the SQWIP active region ($\lambda\gg W\sim 0.1$~$\mu$m), which
corresponds to practical experimental conditions. In this case the Poisson
equation reduces to the one-dimensional equation (along the $x$-axis), which can
be solved analytically to obtain the QW potential:

\begin{equation} \label{pot-eq}
\delta\varphi(y,t) = -\frac{e\WW}{\eps\eps_0}\, \delta \Sigma(y,t) .
\end{equation}

\noindent Here $\WW=W_e W_c/W$ is the reduced SQWIP thickness ($W_e$ and
$W_c$ are the effective emitter and collector barrier
thicknesses,~\cite{ErshovAPLN} respectively, and $W=W_e+W_c$), and $\eps\eps_0$
is the dielectric permittivity. We assumed that the contacts are highly
conductive, i.~e. respond instantaneously to the change of the QW charge. The
injection current density is given by $\delta j_e=\gamma_e\, \delta E_e =
\gamma_e\,
\delta \varphi /W_e$, where $\gamma_e=dj_e/dE_e$ is the differential conductivity
of the injection barrier. The total photocurrent density in SQWIP is given by
the formula:

\begin{eqnarray} \label{cur-eq}
\delta j(y,t)&=& e\sigma\Sigma \delta I \left[\zeta
\frac{W_c}{W} -(1-\zeta)\frac{W_e}{W}\right] + \nonumber \\
&& \delta j_e \left[ \beta + (1-\beta) \frac{W_e}{W} \right],
\end{eqnarray}

\noindent where $\zeta$ is the probability for the photoexcited electron to
reach the collector ($(1-\zeta)$ is the probability to reach the emitter). The
first term in eq.~(\ref{cur-eq}) corresponds to the current induced by the
electrons excited from the QW (primary photocurrent), and the second term
corresponds to the current due to the electrons injected from the emitter
(secondary or multiplied photocurrent).

The solution of eqs.~(\ref{cont-eq}--\ref{cur-eq}) for the case of infrared
radiation modulated harmonically in space and time $\delta I(y,t)=\delta I
\exp[i(ky-\omega t)]$ is given by $\delta j(y,t)=\delta j(k,\omega)
\exp[i(ky-\omega t)]=g(k,\omega) \,e\,\sigma \Sigma\, \delta I(y,t)$, where
the photocurrent gain $g$ is expressed as:

\begin{eqnarray} \label{gain}
g(k,\omega)&=&\left[\zeta-\frac{W_e}{W}\right] + \nonumber \\ &&\left[
\frac{\beta}{1-\beta} + \frac{W_e}{W}\right] \times \frac{1}{1-i\omega \tau +
\lambda^2 k^2 } .
\end{eqnarray}

\noindent The characteristic time constant $\tau$ and length $\lambda$ are
given by the formulas:

\begin{equation} \label{tau}
\tau=(\eps \eps_0 W)/\left[ (1-\beta)\gamma_e\, W_c \right],
\end{equation}

\begin{equation} \label{lambda} \lambda = \sqrt{[e\mu\Sigma
\WW/(\varepsilon\varepsilon_0) +D]\tau} \approx \sqrt{v \WW \tau} ,
\end{equation}

\noindent where $v=\sigma_{2D}/(\varepsilon\varepsilon_0)$ is the velocity
of charge relaxation in 2DEG with conductivity
$\sigma_{2D}=e\mu\Sigma$.~\cite{Shik2D} Formula~(\ref{gain}) generalizes a
previously obtained formula for the photocurrent gain in the 1D case
($k=0$).~\cite{ErshovAPLN} The lower limit of the gain $g=\zeta-W_e/W$ is
determined by the primary photocurrent. The dispersion of the gain is due to the
cut-off of the injection from the emitter, which is caused by the ``freezing''
of the QW recharging processes at high frequencies $\omega \gg 1/\tau$, or
lateral smearing of the photoinduced charge at high spatial frequencies $k\gg
1/\lambda$. The characteristic time constant $\tau$ is the QW recharging time,
or the time of establishing equilibrium at the injecting contact. It depends
strongly on operating conditions and SQWIP design, and its value can span over a
wide range $\sim$~10$^{-9}$--10$^{-3}$~s.~\cite{ErshovIEEE96} It should be noted
that the time constant $\tau$ determines the frequency dispersion of various
SQWIP characteristics -- admittance, noise current, photocurrent,
etc.~\cite{ErshovIEEE96,ErshovAPLN} The characteristic length $\lambda$ can be
interpreted as an effective diffusion length $\sqrt{\DD \tau}$, where $\tau$ can
be considered as the life-time of the non-equilibrium electrons in the QW, and
$\DD=v\WW$ is an effective diffusion coefficient. However, this effective
diffusion is different from the real diffusion caused by the gradient of carrier
concentration. The lateral out-diffusion is related to the drift of the QW
electrons in the in-plane field created by the non-uniform distribution of the
QW charge and charges induced on the emitter and collector contacts. The
effective diffusion coefficient can also be represented as $\DD=1/(RC)$, where
$R$ is the sheet resistivity of the 2DEG, and $C=\eps\eps_0/\WW$ is the
effective SQWIP capacitance (compare with ref.~\onlinecite{Livescu89}). An
estimate of $\DD$ for typical SQWIP structures ($W_e=W_c=500~\AA$, $\Sigma\sim$
10$^{11}$--10$^{12}$~cm$^{-2}$, $\mu\sim$10$^3$~cm$^2$/Vs) gives $\DD\sim
10^2$--10$^3$~cm$^2$/s, which is much larger than the typical values of
diffusion coefficient $D\sim 10$~cm$^2$/s. An estimate of the characteristic
length gives $\lambda\sim$10$^1$--10$^4$~$\mu$m ($\lambda\sim$100~$\mu$m for
SQWIP presented in ref.~\onlinecite{BandaraS93}), which can be comparable with
or larger than the lateral dimensions of SQWIPs.

Let us now consider a spatial resolution of SQWIP when it is used in QWIP-LED
imaging device. To characterize the spatial resolution of QWIP-LED, we calculate
the modulation transfer function~\cite{VincentBook} (MTF) of SQWIP, which is the
ratio of the amplitude of the small-signal photocurrent for modulated
illumination to that for the uniform steady-state illumination:

\begin{eqnarray} \label{MTF}
{\cal M}(\omega,k) &\equiv & \frac{\delta j(\omega,k)}{\delta j(0,0)}=
\left[\left( \zeta-\frac{W_e}{W}\right)+
\left(\frac{\beta}{1-\beta} +\frac{W_e}{W}\right)\times \right. \nonumber \\
&&\left. \frac{1}{1-i\omega\tau +\lambda^2 k^2} \right]
\left(\frac{\beta} {1-\beta} +\zeta\right)^{-1} .  \end{eqnarray}

\noindent For SQWIPs with large photocurrent gain ($\beta\rightarrow$1) this
formula reduces to ${\cal M}=(1-i\omega\tau +\lambda^2 k^2)^{-1}$, so that
MTF is strongly degraded if $\omega \gg 1/\tau$ or $k\gg 1/\lambda$. On the
other hand, in SQWIPs with $\beta=0$ (this case is realized, for example,
for SQWIP with tunneling injection from emitter to the QW~\cite{BandaraST93})
MTF approaches a value of ${\cal M}=1-W_e/(\zeta W)$ in the limit of
high-frequency time or spatial modulation. Theoretically, one can expect
very large (negative) values of MTF in the case $\zeta \ll 1$.

If the characteristic spreading length $\lambda$ exceeds the lateral SQWIP
dimension, the variation of the electron density in the QW is spread uniformly
over the whole SQWIP area under non-uniform illumination conditions. In this
case all physical processes are essentially one-dimensional. This effect can
play both positive and negative role for QWIP operation. Under localized
excitation, the whole QWIP area work as a photosensitive area. This effect
results in lower photocurrent densities and helps to avoid undesirable effect of
responsivity nonlinearity at high excitation power.~\cite{ErshovNonlAPL} On the
other hand, if large-area SQWIP has only one defect resulting in the
QW-collector leakage, the whole SQWIP will be shortened down and thus defective.
For SQWIP operation under low-power illumination and electrical read-out of
output signal, the photocurrent spreading effects are not important, since the
total photocurrent is integrated over the SQWIP area.

Lateral spreading of the photoinduced charge in the QWs and resulting
photocurrent spreading can limit the spatial resolution and even kill the
imaging capabilities of pixelless QWIP-LED devices. The guidelines for
decreasing spreading length $\lambda$ and, therefore, for improving spatial
QWIP-LED resolution, are obvious from eq.~(\ref{lambda}). One possible solution
is to decrease $\tau$ by designing SQWIP with very high differential
conductivity of the injection barrier $\gamma_e$. Another way is to decrease the
in-plane conductivity (mobility) of carriers in the QW. This can be achieved,
for example, by periodic modification of the QW structure (e. g. by QW
intermixing), or by using an array of weakly coupled quantum dots instead of
QW.~\cite{QD} The decrease of the carrier mobility can be also achieved by using
$p$-type SQWIP ($p$-type doped QW), since the hole mobility is typically much
lower than the electron mobility (in GaAs/AlGaAs system).

In conclusion, we considered physical effects responsible for lateral
photocurrent spreading in SQWIPs. Analytical expressions for the characteristic
spreading length, photocurrent gain, and MTF were obtained.

The author thank Dr.~H.~C.~Liu for valuable discussions and comments on the
manuscript. The work has been partially supported by Electronic Communication
Frontier Research and Development Grant of the Ministry of Post and
Telecommunications, Japan, and Research Fund of the University of Aizu.

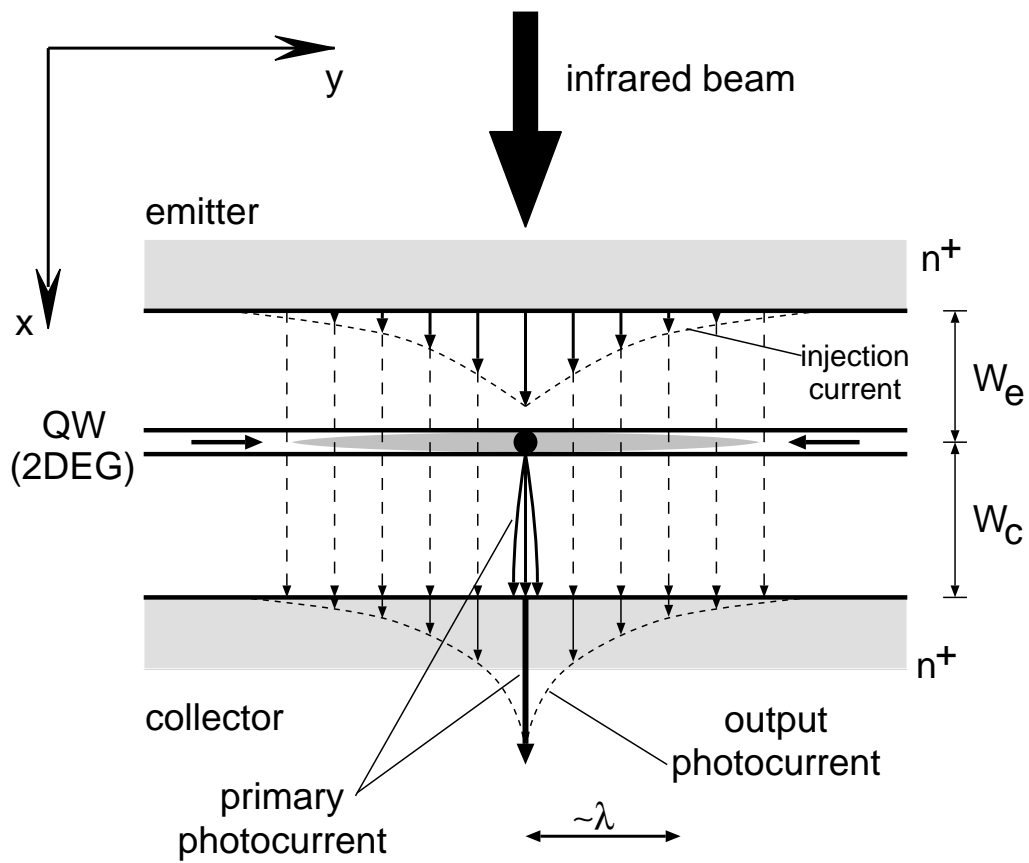
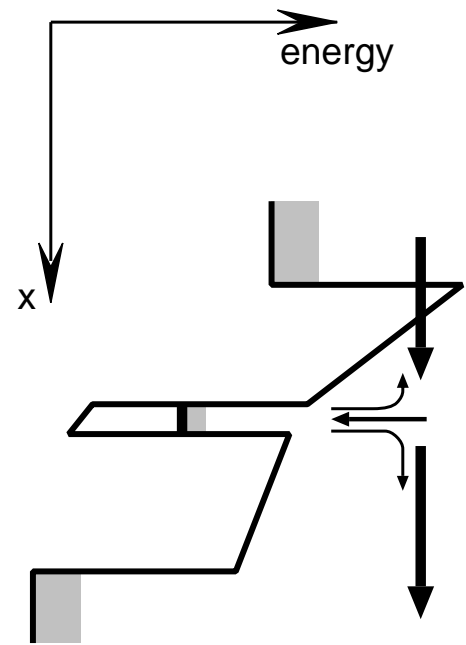
\begin{figure}
\caption{\label{fig:1}
(a) Structure and physical effects in SQWIP under localized excitation and (b)
conduction band diagram of SQWIP. }
\end{figure}

\end{document}